\documentclass[a4paper,aip,twocolumn,10pt,superscriptaddress]{revtex4-1}
\usepackage{textcomp}
\usepackage{graphicx}
\usepackage{xr}
\usepackage{amsmath}
\usepackage{dcolumn}
\usepackage{bm}
\usepackage{mathtools}
\usepackage{soul}
\usepackage[utf8]{inputenc}
\usepackage[T1]{fontenc}
\usepackage{mathptmx}
\usepackage{etoolbox}

\usepackage{color}
\usepackage{xcolor}
\usepackage{soul}

\begin{document}

\title{Excitons, Optical Spectra, and Electronic Properties of Semiconducting Hf-based MXenes}
\author{Nilesh Kumar}
\affiliation{Department of Physics, Faculty of Science, University of Ostrava, 30.~dubna 22, 701 03 Ostrava, Czech Republic}
\author{Miroslav Kolos}
\affiliation{Department of Physics, Faculty of Science, University of Ostrava, 30.~dubna 22, 701 03 Ostrava, Czech Republic}
\author{Sitangshu Bhattacharya}
\email{sitangshu@iiita.ac.in}
\affiliation{Electronic Structure Theory Group, Department of Electronics and Communication Engineering, Indian Institute of Information Technology-Allahabad, Uttar Pradesh 211015, India}
\author{Franti\v{s}ek Karlick\'{y}}
\email{frantisek.karlicky@osu.cz}
\affiliation{Department of Physics, Faculty of Science, University of Ostrava, 30.~dubna 22, 701 03 Ostrava, Czech Republic}

\begin{abstract}
Semiconducting MXenes are an intriguing two-dimensional (2D) material class with promising electronic and optoelectronic properties. 
Here, we focused on recently prepared Hf-based MXenes, namely Hf$_3$C$_2$O$_2$ and Hf$_2$CO$_2$. 
Using the first-principles calculation and excited state corrections, we proved its dynamical stability, reconciled its semiconducting behavior, and obtained fundamental gaps by the many-body GW method (indirect 1.1~eV and 2.2~eV, respectively, direct 1.4~eV and 3.5~eV, respectively).
Using the Bethe-Salpeter equation (BSE) we subsequently provided optical gaps (0.9~eV and 2.7~eV, respectively), exciton binding energies, absorption spectra, and other properties of excitons in both Hf-based MXenes. 
The indirect character of both 2D materials further allowed a significant decrease of excitation energies by considering indirect excitons with exciton momentum along the $\Gamma$-M path in the Brillouin zone. 
The first bright excitons are strongly delocalized in real space while contributed by only a limited number of electron-hole pairs around the M point in the k-space from the valence and conduction band.
A diverse range of excitonic states in Hf$_3$C$_2$O$_2$ MXene lead to a 4\% and 13\% absorptance for the first and second peaks in the infrared region of absorption spectra, respectively. In contrast, a prominent 28\% absorptance peak in the visible region appears in Hf$_2$CO$_2$ MXene. 
Results from radiative lifetime calculations indicate the promising potential of these materials in optoelectric devices requiring sustained and efficient exciton behavior. 
\end{abstract}
\maketitle

\section{Introduction}
\label{intro}
In recent years, the exploration of two-dimensional (2D) monolayers and their derivatives has captivated the attention of researchers due to their remarkable potential in the realm of electrical and optoelectronic applications. 
The groundbreaking properties exhibited by graphene have inspired scientific inquiry into novel 2D materials possessing finite bandgaps. 
Among these, 2D MXenes have emerged as a distinct class of materials exhibiting extraordinary characteristics and versatile applications in nanotechnology.\cite{naguib201425th,anasori20172d,gogotsi2019rise} 
MXenes are a unique and intriguing material class belonging to the family of 2D transition metal carbides, nitrides, or carbonitrides. MXenes is M$_{n+1}$X$_n$T$_x$, where M is a transition metal (such as titanium, scandium, or hafnium), X is carbon and/or nitrogen, $n$ is the number of layers ($n \leq$ 4), and T denotes surface functional groups. 

MXenes were first discovered in 2011 by Gogotsi et al. (Ti$_3$C$_2$T$_x$), and since more than 20 MXenes have been synthesized up to now.\cite{naguib2011two}
Following these experimental syntheses, many other MXene compositions have been predicted theoretically. \cite{lim2022fundamentals,anasori2022mxenes} 
2D semiconducting MXenes have a tunable band gap that can be adjusted by changing the transition metal, surface functionalization (T = O, F, OH, Cl, ...), strain, defects, or intercalated species.\cite{hart2019control, Ketolainen2022, Sakhraoui2022, kumar2023oxygen} 
This tunability enables customized band topologies, making semiconducting MXenes highly valuable for various electronic device applications.\cite{Kim2020, Champagne2021} 

Experimental preparation of MXenes is often realized from the corresponding layered materials, MAX phases, where A is an A group (mostly IIIA and IVA) element. 
Al-containing MAX phases Hf$_2$AlC and Hf$_3$AlC$_2$ were experimentally synthesized too.\cite{Lapauw2016} 
However, the subsequent preparation of Hf-containing MXene from MAX phase was not yet done and the transition metal Hf was identified as more inclined to form a different family of layered ternary and quaternary transition metal carbides beyond the MAX phases (as Hf$_3$Al$_3$C$_5$, Hf$_3$Al$_4$C$_6$, Hf$_2$Al$_4$C$_5$, or Hf$_2$[Al(Si)]$_4$C$_5$).\cite{Zhou2013}
In 2017, Zhou et al.\cite{Zhou2017} experimentally prepared a 2D Hf-based MXenes Hf$_
3$C$_2$T$_x$ by selective etching
of a layered parent Hf$_3$[Al(Si)]$_4$C$_6$ compound. 
Such MXenes were determined to be flexible, conductive, and a good candidate for anode material for metal-ion intercalation. 
Further, theoretical findings of the Ref. \cite{Zhou2017} predicted the metallic behavior with good electrical conductivity for Hf$_3$C$_2$T$_2$ MXenes with T = O, F, and OH, particularly highlighting the robustness of Hf$_3$C$_2$O$_2$ with a superior mechanical strength of 417 GPa value of elastic constant. 
In addition, Zha et al.\cite{Zha2017} claimed using the density functional theory (DFT) calculations Hf$_3$C$_2$O$_2$ MXene as a semi-metal or conductor (depending on DFT level). 
Finally, Hf$_3$C$_2$O$_2$ was also suggested to be a promising material for electronic devices modulated by strain because DFT calculations predicted a transition between semi-metal and semi-conductor induced by strain.\cite{zha2016thermal} 
On the other hand, the moderate and tunable gap of MXene Hf$_2$CO$_2$ made itself a unique 2D semiconductor.\cite{Khazaei2013, zhang2019prediction, Champagne2021}

In the realm of previous computational studies, there are various investigations on the electronic and optical properties of MXenes, which have primarily relied on generalized-gradient approximation (GGA) to density functional theory (DFT) and other higher levels of density functionals (as hybrids). 
To reconcile unclarities mainly in Hf$_3$C$_2$O$_2$ electronic properties (we finally assigned here the material as a semiconductor, analogically to Ti$_3$C$_2$O$_2$ case\cite{kumar2023oxygen}), there is a necessity to study the material in more detail. 
DFT is generally suitable to study the structure, the trends in composition, or the impact of strain in 2D materials.\cite{kumar2021strain, Kumar2022} 
However, the reason for going beyond DFT is that DFT does not inherently include the effects of excitons (electron-hole pairs), which are crucial in understanding the optical properties of 2D materials.\cite{bernardi2013extraordinary,qiu2013optical} 
To overcome these limitations and obtain more accurate electronic and optical properties, it's necessary to consider more advanced many-body methods. 
Many-body perturbation theory, GW approximation for the self-energy calculation\cite{hedin1965new} and subsequent GW+BSE (Bethe-Salpeter equation)\cite{salpeter1951relativistic,fetter2012quantum} including excitonic effects, is a preferred computational method, recently shown as precise-performing for different MXene in comparison to another independent and demanding stochastic many-body method.\cite{Dubecky2023} 

In this study, we focus our attention on semiconducting oxygen-terminated Hf-based MXenes and we carefully describe them using the many-body perturbation theory. 
Both Hf-based MXenes, Hf$_3$C$_2$O$_2$ and Hf$_2$CO$_2$, contain heavy 5d elements, so it was also demanded to include the spin-orbit coupling (SOC) in the calculations. 
We present its accurate electronic and optical properties (including absorptance spectra and a radiative lifetime of exciton) and we describe their indirect nature. 

\section{Computational Methods}
\label{methods}
For ground-state calculations, geometrical conformers of 2D-MXenes were prepared following our previous work on Ti$_3$C$_2$O$_2$ case\cite{kumar2023oxygen} (see a set of highly symmetric structures in Figures S1 and S2 in the \textcolor{blue!80!black}{supplementary material}). 
Vacuum space was added along the out-of-plane axis (z-direction), leading to $c$ lattice constant of 22.5~{\AA} and 19.6~{\AA} for Hf$_3$C$_2$O$_2$ and Hf$_2$CO$_2$, respectively, preventing interlayer interactions. 
Density functional theory (DFT) calculations were performed using the Quantum ESPRESSO (QE) package,\cite{giannozzi2017advanced} utilizing a fully relativistic norm-conserving pseudopotential\cite{hamann2013optimized} enabling spin-orbit coupling (SOC). 
A Perdew-Burke-Ernzerhof (PBE) exchange-correlation functional\cite{perdew1996generalized} within the generalized gradient approximation (GGA) to DFT was used, and a kinetic cutoff energy of 70 Ry ensured accurate ground-state calculations. 
The convergence behavior of the kinetic energy cutoff is plotted in Figure S3 in \textcolor{blue!80!black}{supplementary material}. 
A set of 24$\times$24$\times$1 k-mesh grid was used for the structure relaxation without using any constraint through the force and total energy minimization with respect to all atomic positions and unit cell parameters. 
A self-consistency convergence threshold of 1~$\times$ 10$^{-8}$~Ry was used in the calculations. 
Self-consistent and non-self-consistent calculations were performed, incorporating non-collinear spin-orbit interactions. 
To confirm the dynamic stability of both Hf-based MXenes, phonon band dispersion calculations were performed using density functional perturbation theory (DFPT) on a 10$\times$10$\times$1 \textbf{q}-grid and a threshold 1~$\times$ 10$^{-15}$~Ry. 
Non-analytical part (using dielectric matrix and Born effective charges) was included.

Excited state corrections were computed via the many-body perturbation theory (MBPT) YAMBO code (version 5.1.0),\cite{marini2009yambo,sangalli2019many} employing the GW approximation to include quasiparticle self-energy corrections. A converged value of 70~Ry of cutoff energy is used in the sum of the exchange self-energy. 
The converged value 14~Ry of screening cutoff energy is used to calculate the polarization function in GW calculations. A total of 300 polarization bands were utilized, comprising 56 bands for Hf$_3$C$_2$O$_2$ and 40 bands for Hf$_2$CO$_2$, all of which represent the filled electronic states up to the Fermi level in the respective materials. These bands were employed to calculate the irreducible polarization response function. 
The GW calculations utilized a random integration method\cite{pulci1998ab} with 1 $\times$ 10$^{6}$ \textbf{q} points and a truncated Coulomb potential.\cite{rozzi2006exact, castro2009exact} Box-like Coulomb cutoff geometries of 21.98~Å and 19.09~Å were applied in the z-direction to the sheet for Hf$_3$C$_2$O$_2$ and Hf$_2$CO$_2$, respectively. Utilizing Kohn-Sham wavefunctions and GW quasiparticle energies, we computed the linear response optical spectra employing the time-independent Bethe-Salpeter equation (BSE). The exchanged and screened cutoffs are used at the converged value of 30~Ry and 3~Ry to build up the exchange electron-hole attractive and repulsive kernels in the BSE matrix.
Tamm-Dancoff approximation is used, incorporating the resonant matrix element into the BSE Hamiltonian. The BSE matrix was solved using the diagonalization method, where the obtained poles corresponded to optical transition energies. The BSE can be written as an eigenvalue problem for insulating materials possessing occupied valence bands (v) and entirely unoccupied conduction bands (c)\cite{albrecht1998ab, Ketolainen2020}
\begin{equation}
\label{BSE}
\begin{multlined}
(\epsilon_{c\bm{k}}^{\mathrm{GW}}-\epsilon_{v\bm{k}}^{\mathrm{GW}})A_{cv \bm{k}}^\lambda
+\sum_{c'v'\bm{k}'}[2\langle \phi_{c\bm{k}}\phi_{v\bm{k}} \lvert \nu \rvert \phi_{c'\bm{k}'}\phi_{v'\bm{k}'}\rangle \\ 
-\langle \phi_{c\bm{k}}\phi_{c'\bm{k}'} \lvert W \rvert \phi_{v\bm{k}}\phi_{v'\bm{k}'}\rangle
]A_{c'v'\bm{k'}}^\lambda
= E_\lambda A_{cv\bm{k}}^\lambda ,
\end{multlined}
\end{equation}
where $\nu$ is the Coulomb kernel, $1/\lvert \bm{r}-\bm{r}'\rvert$, the eigenvectors $A_{cv \bm{k}}^\lambda$ correspond to the amplitudes of free electron-hole pair configurations composed of electron states $\rvert \phi_{c\bm{k}}\rangle$ and hole states $\rvert \phi_{v\bm{k}}\rangle$ and the eigenenergies $E_\lambda$ correspond to the excitation energies (with optical gap $\Delta_\mathrm{opt} \equiv E_\lambda$, for $\lambda$ from first nonzero transition, i.e., first bright exciton). 
The exciton binding energy was finally estimated as $E_\mathrm{b}=\Delta_{\mathrm{GW,dir}}-\Delta_{\mathrm{opt,dir}}$, where "dir" index denotes direct gap.

Indeed, in our GW+BSE calculations employing a truncated Coulomb potential, the spectra were obtained in terms of macroscopic polarizability $\alpha$, a well-defined quantity intimately linked to absorption spectra. For a 2D system, {$\alpha$} is expressed in units of length 
and is defined as\cite{rasmussen2016efficient,guandalini2023efficient}
\begin{equation}
\label{Alpha}
\alpha(\omega) = -{\lim_{\textbf{q}\to 0}} {\frac{L}{4 {\pi}q^{2}}} {\chi_{00}}(\textbf{q},{\omega}) ,
\end{equation}
where $\chi_{00}(\textbf{q},{\omega})$ is the non-interacting density response function or irreducible polarizability, and $L$ is the length of the computational cell in the $z$-direction.\\
The dielectric function ({$\epsilon$}) is extracted from this macroscopic polarizability ($\alpha=\alpha_1 + i\alpha_2$) using the equation\cite{molina2020magneto}
\begin{equation}
\label{Epsilon}
     \epsilon(E) = (1+4 {\pi} {\frac{\alpha_1}{L}}) + i4 {\pi} {\frac{\alpha_2}{L}}
\end{equation}
and $\epsilon(E)=\epsilon_{1}(E)+i\epsilon_{2}(E)$, where  $\epsilon_{1}(E)$ and $\epsilon_{2}(E)$ are the real and imaginary part of the dielectric function. The absorptance is calculated using the relation\cite{Ketolainen2020} 
$A(E) = 1-\mathrm{exp}[-\epsilon {_2}E L/{\hbar}c]$, where $E$ is the energy of incoming photon, $\hbar$ is reduced Planck's constant, and $c$ is speed of light. 

Additionally, we determine the exciton radiative lifetime $ \tau_{\lambda}^{0}$ utilizing the relation (in SI units)\cite{chen2018theory}
\begin{equation}
\label{radiative_lifetime}
    \tau_{\lambda}^{0} = \frac{\epsilon_0\hbar^2 c}{e^2 E_\lambda} \frac{A_{uc}}{\mu^{2}_{\lambda}} ,
\end{equation}
where $A_{uc}$ is the area of the unit cell, $\epsilon_0$ is vacuum permittivity, $E_\lambda$ is the exciton energy related to Equation \ref{BSE}, $e$ is the elementary charge, and $\mu^2_\lambda$=$\lvert \sum_{cv \bm{k}} A_{cv \bm{k}}^\lambda \langle \phi_{c\bm{k}}|\textbf{r}|\phi_{v\bm{k}} \rangle \rvert^2/N_k$ is $\lambda$-exciton intensity\cite{yambo_mu} given by the linear combination of the square of the transition matrix elements between electron-hole pairs with the excitonic weights $A_{cv \bm{k}}^\lambda$ (from Equation \ref{BSE}) divided by the number of k-points.\cite{palummo2015exciton, chen2018theory,chen2019ab} 
The average radiative lifetime $\langle \tau_{\lambda}^T \rangle$ of an exciton in state $\lambda$ at temperature $T$ is given by  \cite{palummo2015exciton}
\begin{equation}
\label{avg_radiative_lifetime}
\langle \tau_{\lambda}^T \rangle = \frac{3}{4} \tau_{\lambda}^{0}  \frac{2 M_{\lambda} c^2}{E_{\lambda}^2} k_BT ,
\end{equation}
where $M_{\lambda}$ is the exciton mass, $k_B$ is the Boltzmann constant, and zero-momentum approximation $E_\lambda(q)\approx E_\lambda(q=0)$ is used.\cite{chen2018theory} 
\section{Results and Discussion}
\label{results}
\subsection{Structural properties and stability}
\label{structures and stability}
\begin{figure}[!htb]
\centering
\includegraphics[width=8.5cm]{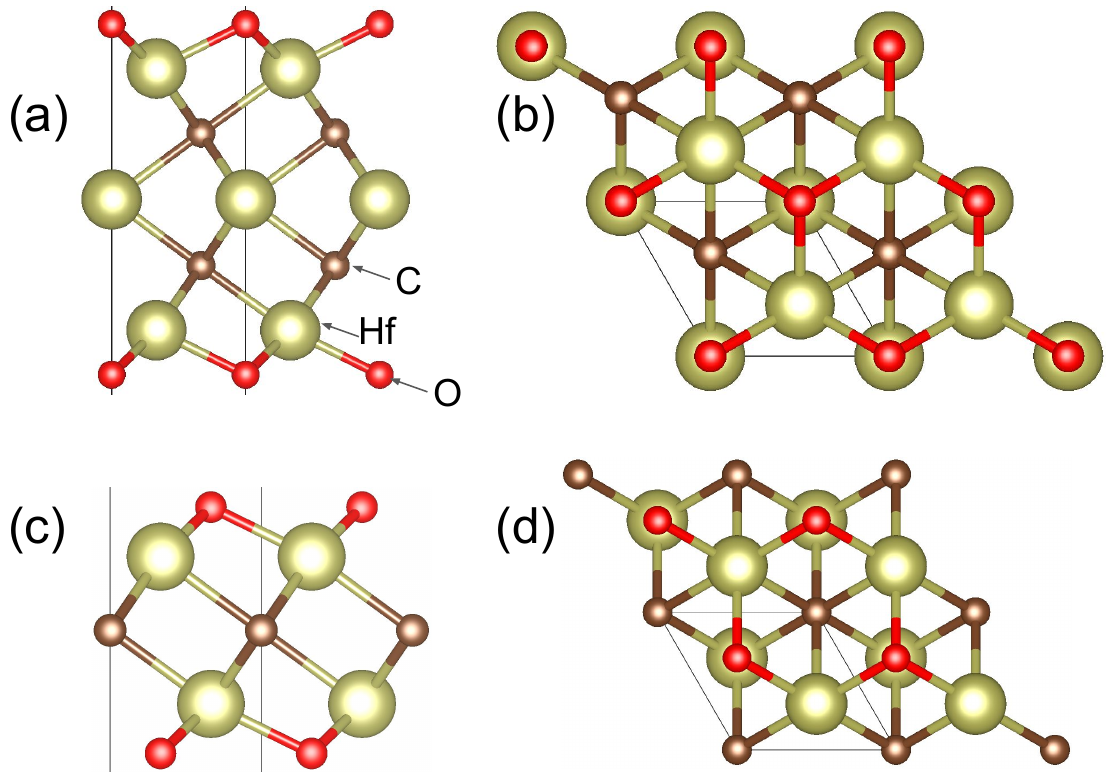}
\caption {\label{fig: structures} (a,c) The side and (b,d) the top view of the energetically most favorable conformers of (a,b) Hf$_3$C$_2$O$_2$ and (c,d) Hf$_2$CO$_2$ MXenes. The black line defines the unit cell.}
\end{figure}

Firstly, we refined the lattice structure of 12 possible high-symmetry conformers of the Hf$_3$C$_2$O$_2$ unit cell and corresponding 8 conformers of the Hf$_2$CO$_2$ unit cell, as these MXenes can have energetically comparable phases.\cite{khazaei2013novel} 
Information about the optimization of all Hf$_3$C$_2$O$_2$ and Hf$_2$CO$_2$ conformers are mentioned in the \textcolor{blue!80!black}{supplementary material} in Tables S1 and S2, respectively, and the optimized structures of the conformers are plotted in Figures S1 and S2, respectively. 
The most stable conformer (A2) of both Hf$_3$C$_2$O$_2$ and Hf$_2$CO$_2$ MXenes contains the oxygen atom at the hollow site. 
The second energetically favorable conformers, i.e., A4 and B2 for Hf$_3$C$_2$O$_2$ and Hf$_2$CO$_2$, respectively, are significantly energetically higher (1.11~eV and 0.64~eV above the lowest conformers A2), and therefore not expected to be present in real samples due to their higher energy state. 
Figure \ref{fig: structures} shows the side and top views of the most stable optimized structures (A2) used in the subsequent calculations. 

Both favorable structures of Hf$_3$C$_2$O$_2$ and Hf$_2$CO$_2$ MXenes exhibit hexagonal configurations.  
Hf$_3$C$_2$O$_2$ structure of Figure \ref{fig: structures} has a D$_{3h}$ (-62m) symmetry while Hf$_2$CO$_2$ has a D$_{3d}$ (-3m) symmetry. 
The primary distinction between D$_{3h}$ and D$_{3d}$ point groups is the presence (or lack) of certain symmetry elements, namely the horizontal mirror plane (h) in  D$_{3h}$ and the vertical mirror plane (d) in D$_{3d}$. 
The optimized in-plane lattice constant is 3.26~{\AA}, the same for both Hf$_3$C$_2$O$_2$ and Hf$_2$CO$_2$ MXenes. 
This implies that these compounds have an equivalent area of the unit cell, which indicates that they may have comparable densities and atomic configurations along the crystallographic directions. 
In Hf$_3$C$_2$O$_2$, the Hf-C bond lengths range from 2.32~{\AA} to 2.34~{\AA}, reflecting the variation in bonding distances within the material. 
Additionally, the Hf-O bond length in Hf$_3$C$_2$O$_2$ is 2.10~Å. While in Hf$_2$CO$_2$, the Hf-C bond length is 2.34~{\AA}, indicating a consistent bonding distance between Hf and C atoms.  
The Hf-O bond length in Hf$_2$CO$_2$ matches that of Hf$_3$C$_2$O$_2$ at 2.10 {\AA}.

\begin{figure}[!htb]
  \centering
\includegraphics[width=8.5cm]{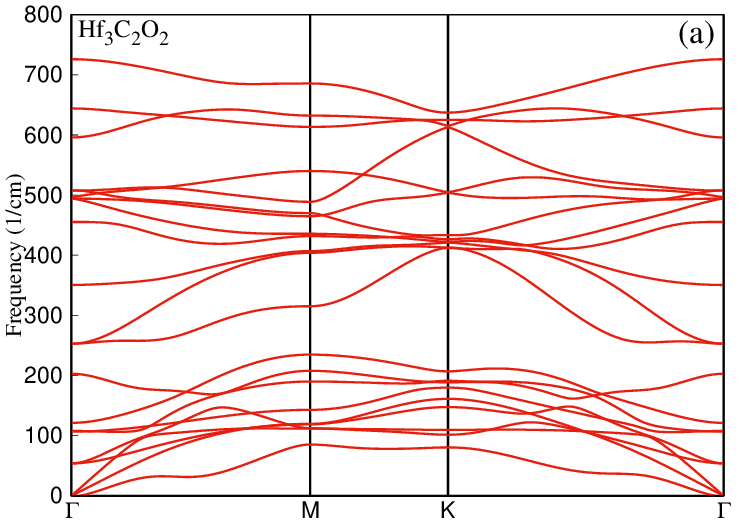}
\includegraphics[width=8.5cm]{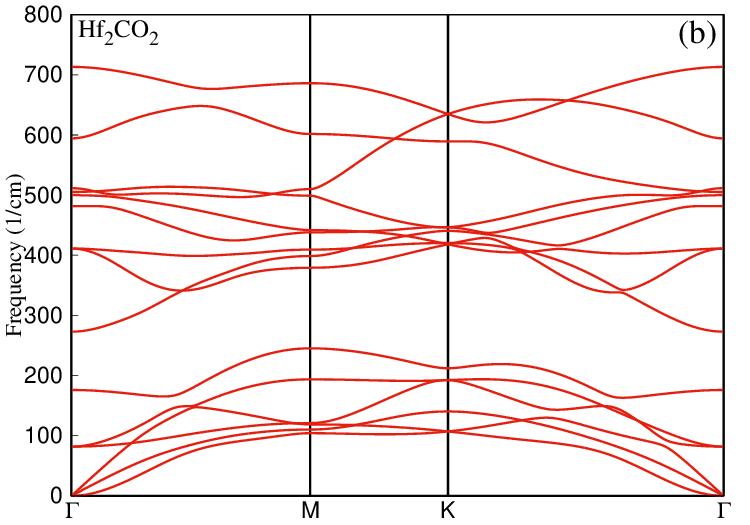}
\caption{\label{fig:bands_QE_phonon}Phonon band structures of (a) Hf$_3$C$_2$O$_2$ and (b) Hf$_2$CO$_2$ MXenes.}
\end{figure}

We also computed the phonon band structures using density functional perturbation theory (DFPT) for both Hf-based MXenes, employing a 24$\times$24$\times$1 k-mesh grid. 
The resulting phonon band structures, depicted in Figure \ref{fig:bands_QE_phonon} reveal positive frequencies throughout the Brillouin zone (BZ).  
This stability signifies that the crystal lattice vibrations are secure and devoid of spontaneous deformations or phase transitions.

\subsection{Electronic band structures and charge analysis from DFT}
\label{dft}
\begin{figure}[!htb]
  \centering
 \includegraphics[width=8.5cm]{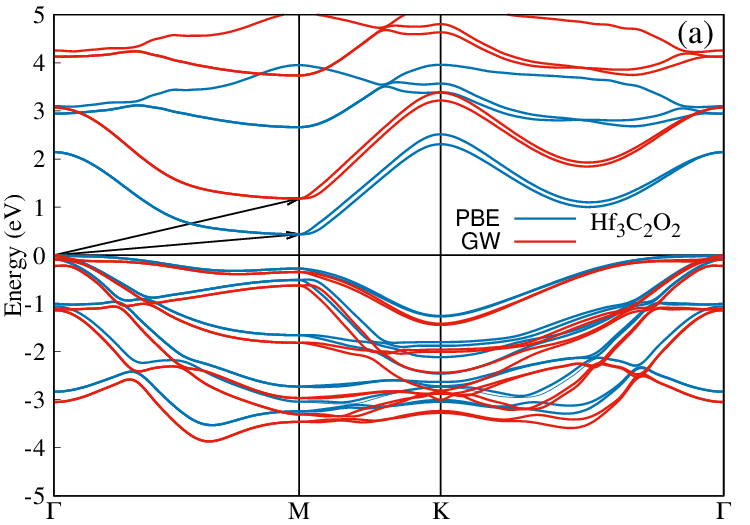}
\includegraphics[width=8.5cm]{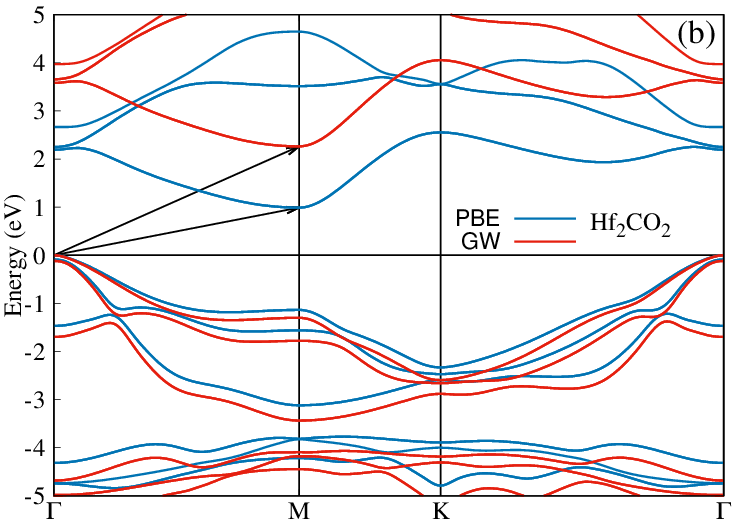}
\caption{\label{fig:bandstructure} DFT (blue lines) and quasiparticle GW (red lines) band structures of (a) Hf$_3$C$_2$O$_2$ and (b) Hf$_2$CO$_2$ MXenes with included spin-orbit coupling (SOC). Black arrows highlight indirect gaps. The Fermi level is set to zero.}
\end{figure}

In our study, we firstly calculated the ground-state DFT electronic band structures of both Hf-based MXenes using 30$\times$30$\times$1 k-mesh grid, plotted in Figure \ref{fig:bandstructure} (blue line). 
Given the substantial presence of hafnium (Hf) atoms, we conducted comprehensive spin-orbit coupling (SOC) calculations to account for relativistic effects. 
The DFT electronic band structures are also calculated without SOC consideration, plotted in Figure S4 of the \textcolor{blue!80!black}{supplementary material}. 
The significant influence of SOC was evident mainly in Hf$_3$C$_2$O$_2$, as indicated by the observed energy band splitting in the electronic band structure around the K point (Figure S4). 
We determined the DFT indirect band gaps of 0.43~eV and 0.99~eV between the high-symmetry points {$\Gamma$} (the top of the valence band) and M (the bottom of the conduction band) for Hf$_3$C$_2$O$_2$ and Hf$_2$CO$_2$, respectively. 
The corresponding direct band gaps of 0.71~eV and 2.12~eV were located in the high-symmetry point M (considering SOC; Figure \ref{fig:bandstructure}). 
The absence of a shift in the positions of the valence band maximum (VBM) and conduction band minimum (CBM) due to SOC is a notable observation in both Hf-based MXenes and indicates that SOC did not significantly affect the band gap in these systems. 
The observed degeneracy in the energy bands at the {$\Gamma$} point implies that specific electronic states at this high-symmetry point have identical energies, even when considering SOC. 
For Hf$_2$CO$_2$ MXene, the DFT results are in good agreement with previously reported values.\cite{zha2016thermal,zhang2019prediction} 
Both Hf-based MXenes exhibited a non-magnetic semiconducting behavior. 

In the case of Hf$_3$C$_2$O$_2$, the Bader charge analysis reveals a notable electron loss in the Hf atoms, with the central atom losing approximately 1.88 electrons and the other two Hf atoms each losing around 2.23 electrons as shown in Table S3 in the \textcolor{blue!80!black}{supplementary material}. This electron deficit indicates the high oxidation states in transition metals within MXene structures. Concurrently, the carbon atoms in Hf$_3$C$_2$O$_2$ each gain about 1.92 electrons. Like the carbon atoms, the terminal oxygen atoms also exhibit an increase in electron density, each gaining about 1.25 electrons, indicating their involvement in substantial bonding interactions with the Hf atoms.

Shifting the focus to Hf$_2$CO$_2$, the Bader charges indicate a similar trend of electron distribution. Here, the Hf atoms lose approximately 2.25 electrons when incorporated into the Hf$_2$CO$_2$ structure, further corroborating their higher oxidation state as detailed in Table S3 in the \textcolor{blue!80!black}{supplementary material}. This electron loss aligns with the behavior typically observed in transition metals in MXene materials. The carbon atom gains about 2 electrons in this structure, pointing to a reduced state. The terminal oxygen atoms in Hf$_2$CO$_2$, like their counterparts in Hf$_3$C$_2$O$_2$, each gain roughly 1.25 electrons.

\subsection{Electronic and optical properties from many-body methods}
\begin{figure}[!htb]
  \centering
\includegraphics[width=8.5cm]{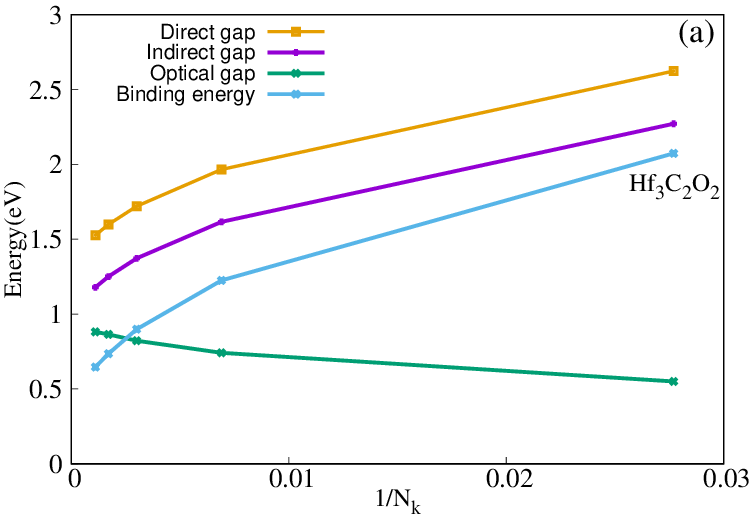}
\includegraphics[width=8.5cm]{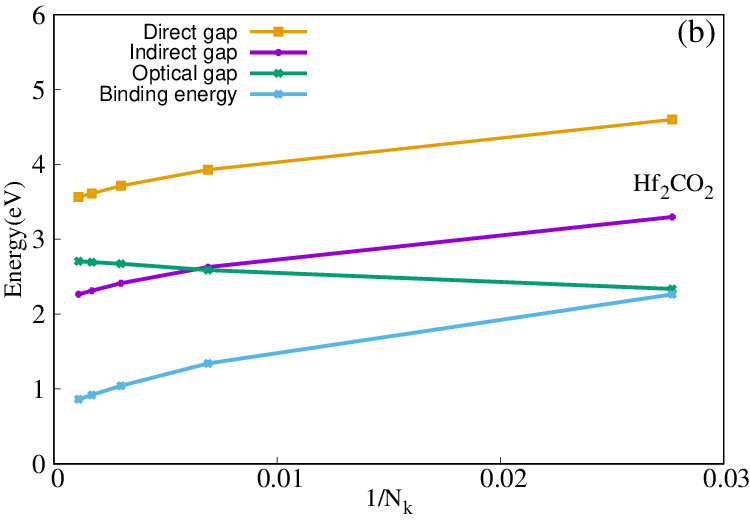}
 \caption{\label{fig:convergence_gap} Convergence behavior of quasiparticle gap, optical gap and binding energy with k-mesh grid $({N_{k}}= {N_{k_{x}}}\times{N_{k_{y}}}\times{N_{k_{z}}}$) in the case of (a) Hf$_3$C$_2$O$_2$ and (b) Hf$_2$CO$_2$ MXenes.}
\end{figure}

To address the issue of inaccuracy in the results of standard DFT calculations and obtain precise results for the electronic band gap and other material properties, employing the many-body methods GW becomes imperative. 
We included the quasi-particle (QP) self-energy corrections on top of the ground-state DFT calculations. 
Meticulous convergence of computational parameters, including cutoff energies, screening cutoff, polarization bands, G-vector grid, and k-mesh grid was paramount to ensure the accuracy and reliability of electronic properties in GW calculations (Figures S5-S7). 
Long-range interactions, the presence of Rydberg states, complex dielectric screening, discretization errors, and the need to balance between localized and delocalized electronic states all contribute to the slow convergence of the fundamental quasi-particle band gaps in 2D materials with a truncated Coulomb potential.\cite{rasmussen2016efficient} 
We also obtained the convergence of the fundamental quasi-particle band gaps with different k-mesh grids ($N_k$), plotted in Figure \ref{fig:convergence_gap}.
The information regarding the convergence of the fundamental quasi-particle gap (direct gap $\Delta_\mathrm{GW,dir}$ and indirect gap $\Delta_\mathrm{GW,ind}$) with the k-mesh grid for MXenes Hf$_3$C$_2$O$_2$ and  Hf$_2$CO$_2$ are mentioned in Table S4 in the \textcolor{blue!80!black}{supplementary material} too. 
We see slow convergence of electronic band gaps concerning the k-point grid, especially for Hf$_3$C$_2$O$_2$, with the same behavior for the optical gap. 
The converged values of all fundamental properties are not easy to obtain even with 30$\times$30$\times$1 k-mesh grid. 
To ensure more precise gaps, we extrapolated the values from the best k-grids,\cite{Dubecky2023} see also Figure S8 in the \textcolor{blue!80!black}{supplementary material} for details (this procedure corrected gaps for $\le$~0.11~eV).
A similar trend (slow gap convergence with $N_k$) was observed for BAs and BP monolayers\cite{Kolos2022} or in Sc$_2$C(OH)$_2$ MXene\cite{Dubecky2023} and was associated with the delocalization of first bound exciton. A detailed discussion on this topic is presented in the subsequent sections of this paper.

The calculated GW band structures with SOC are plotted in Figure \ref{fig:bandstructure}. 
We obtained an indirect band gap of 1.08~eV and a direct band gap of 1.42~eV for Hf$_3$C$_2$O$_2$ MXene, while we obtained an indirect band gap of 2.18~eV and a direct band gap of 3.48~eV for Hf$_2$CO$_2$ MXene (all reported values are extrapolated ones, following Figure S8). 
The GW indirect band gap is located between {$\Gamma$} and M high-symmetry points, while the minimum direct band gap is noticed at high-symmetry point M.

Using BSE (Eqn. \ref{BSE}) on the top of the GW method, we calculated the optical gap $\Delta_\mathrm{opt}$, which corresponds to the position of the first bright excitonic peak, and absorption spectra for both Hf$_3$C$_2$O$_2$ and Hf$_2$CO$_2$ MXenes. 
The convergence of optical gap $\Delta_\mathrm{opt}$ and the binding energy $E_\mathrm{b}$ with the k-mesh grid are also visible in Figure \ref{fig:convergence_gap} and mentioned in Table S4 and Figure S8 in the \textcolor{blue!80!black}{supplementary material}. 
We obtained the converged value of the $\Delta_\mathrm{opt}$ of 0.91~eV {\&} 2.72~eV, and the binding energy $E_\mathrm{b}$ of the first bright exciton of 0.51~eV {\&} 0.76~eV for Hf$_3$C$_2$O$_2$ and Hf$_2$CO$_2$, respectively.

\begin{figure}[!htb]
  \centering
\includegraphics[width=8.35cm]{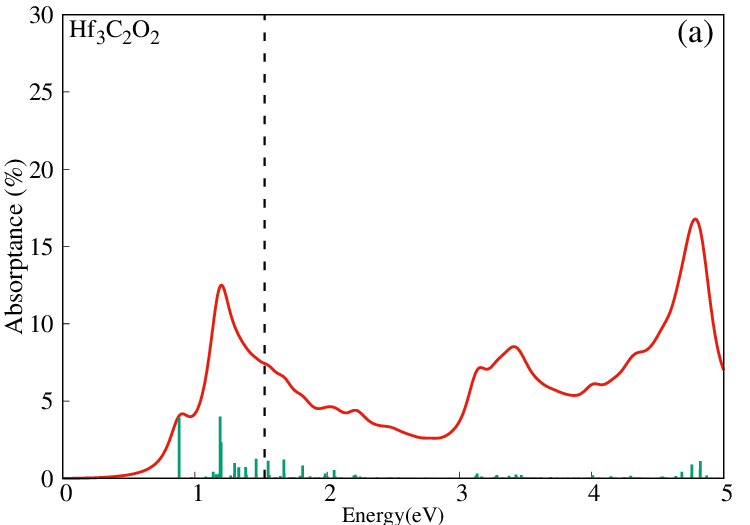}
\includegraphics[width=8.35cm]{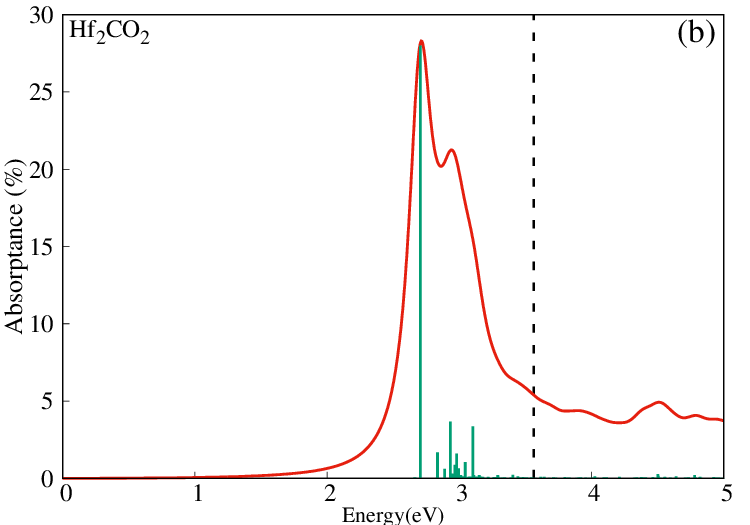}
\caption{\label{fig:absorbance} Optical absorption spectra from BSE for (a) Hf$_3$C$_2$O$_2$ and (b) Hf$_2$CO$_2$ MXenes. The black dotted line denoted the direct quasi-particle gap. Oscillator strengths (arbitrary units) are added as green columns.}
\end{figure}
In our investigation of optical absorption spectra, a meticulous convergence of various calculation parameters was crucial for accurate results. Notably, the choice of the k-mesh grid, the number of bands included in BSE, and other parameters in the BSE significantly impacted the precision of the absorption spectra.\cite{kumar2023oxygen,Dubecky2023,Kolos2022} 
The converged values 3~Ry and 30~Ry of screening cutoff and exchange cutoff, respectively, were used to obtain the accurate absorption spectra (see also Figures S9 and S10 in the \textcolor{blue!80!black}{supplementary material}).
The convergence of the imaginary part of the macroscopic polarizability ({$\alpha$}) concerning k-mesh grids is typically crucial in BSE and it is plotted in Figure S11 for both MXenes. 
Finally, the reliability of the considered energy window for the spectra (the imaginary part of {$\alpha$}) is given by the number of occupied and virtual bands (o,v) included in the BSE calculation. 
Based on Figure S12 in the \textcolor{blue!80!black}{supplementary material}, we used the converged values (12,4) and (6,6) of (o,v) bands for Hf$_3$C$_2$O$_2$ and Hf$_2$CO$_2$, respectively.  
 
In our study, we observed a rich array of excitonic states contributing to the absorption spectra of Hf$_3$C$_2$O$_2$ MXene (Fig.  \ref{fig:absorbance}). 
We obtained the absorptance of 4{\%} for the peak corresponding to the first bright optical transition, but specifically, multiple excitons were identified, leading to the absorptance of 13{\%} for the second peak of spectra within the infrared region (0-1.5~eV) of the spectra.
The absorption spectra revealed a prominent peak, located at 4.8~eV in the near UV region with a strikingly high absorptance of approximately 17{\%}. 
In the case of the Hf$_2$CO$_2$ monolayer, the first peak is located in the visible part of the spectra (1.5-3~eV) with maximum absorptance around 28{\%}. 
We identified 5 and 10 dark excitons for Hf$_3$C$_2$O$_2$ and Hf$_2$CO$_2$, respectively, which are located below the prominent first peak in the spectra.
For completeness, imaginary {$\alpha$} are plotted along oscillator strengths in Figure S13 in the \textcolor{blue!80!black}{supplementary material} using 30$\times$30$\times$1 k-mesh grid for both Hf$_3$C$_2$O$_2$ and Hf$_2$CO$_2$. 

\begin{figure}[!htb]
  \centering
\includegraphics[width=8.5cm]{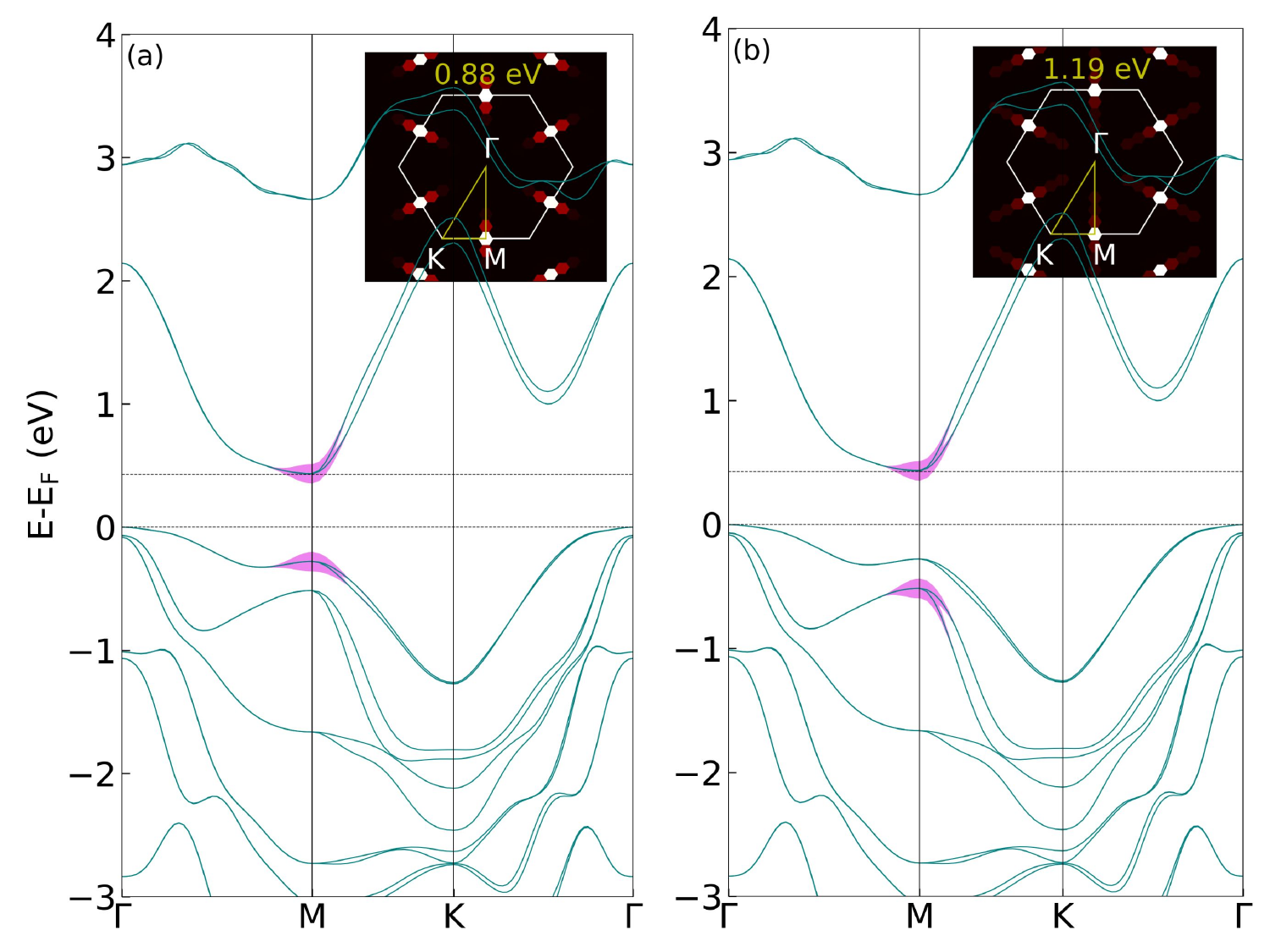}
\includegraphics[width=8.5cm]{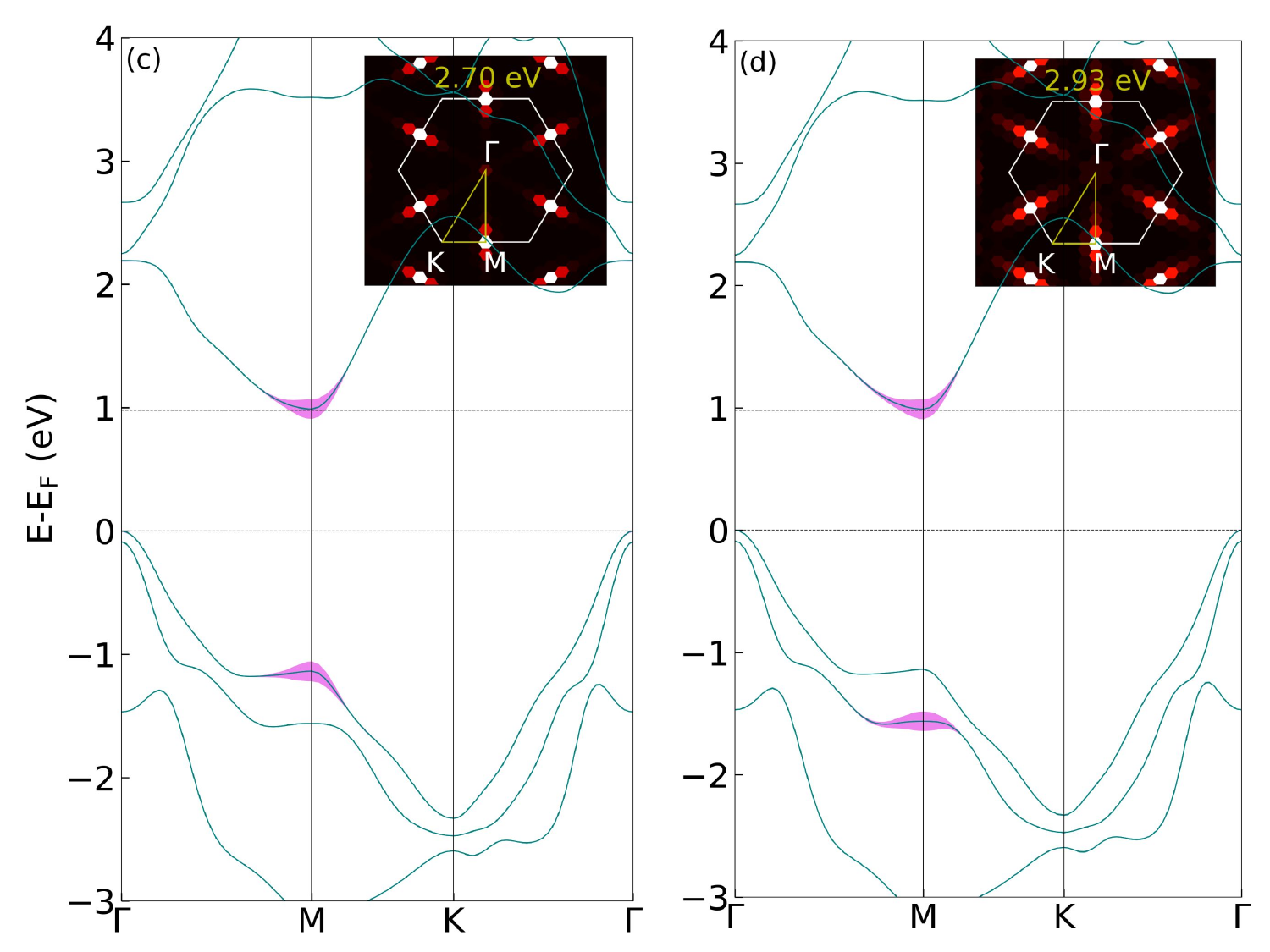}
\caption{\label{fig:exc_weight} Exciton weights in (a-b) Hf$_3$C$_2$O$_2$ and (c-d) Hf$_2$CO$_2$ projected onto the ground-state electronic dispersion. The purple shade (and its vertical thickness) represents the $|A_{cv \bm{k}}^\lambda|$ coefficients (from Eqn. \ref{BSE}) indicating visually which electron–hole pairs contribute to a particular BSE eigenstate: (a) degenerated first-bright exciton with $\lambda$=6-9 and (b) higher exciton $\lambda$= 262, 263 in Hf$_3$C$_2$O$_2$; (c) degenerated first-bright exciton with $\lambda$= 9-12 and (d) higher exciton $\lambda$= 35, 36 in Hf$_2$CO$_2$. The accompanying subplot in each panel displays the exciton weights within the whole Brillouin zone, not resolving particular bands.}
\end{figure}
In our detailed analysis of optical transitions and exciton formation, we investigated the excitonic weights $A_{cv \bm{k}}^\lambda$ (from Equation \ref{BSE}) because the excitonic wave function is expressed in an electron-hole product basis as $\sum_{cv \bm{k}} A_{cv \bm{k}}^\lambda \phi_{c\bm{k}}\phi_{v\bm{k}}$. 
The $A_{cv \bm{k}}^\lambda$ coefficients, therefore, help us to identify which electron-hole pairs in the expansion dominantly contribute to the wave function of the particular exciton (labeled by certain $\lambda$).
The purple shade in Figure \ref{fig:exc_weight} represents $|A_{cv \bm{k}}^\lambda|$ visualizing the important electron-hole contributions labeled by $c,v$ subscripts (band pairs on the y-axis) and by $\bm{k}$ subscript describing the x-axis (missing purple shade means negligible coefficients are not visible in the graph). 
For the first bright excitons in both Hf$_3$C$_2$O$_2$ (Figure \ref{fig:exc_weight}a) and Hf$_2$CO$_2$ (Figure \ref{fig:exc_weight}c), which are consisting from several BSE states $\lambda$ = 6-9 ($E^{6-9}=0.91$ eV) and $\lambda$ = 9-12 ($E^{9-12}=2.72$ eV), respectively, only the valence band and conduction band are important. 
Moreover, only the area around the high-symmetry M point in the Brillouin zone contributes (visible also in the subplots of Figure \ref{fig:exc_weight}). 
Further, we visualized excitons dominantly responsible for the second significant peaks in the spectra of Figure \ref{fig:absorbance}. 
In Hf$_3$C$_2$O$_2$, the exciton at 1.2~eV (Figure \ref{fig:exc_weight}b) is composed mainly of M localized pairs of the conduction band and second valence (VB-1) band. 
The properties of the bright Hf$_2$CO$_2$ exciton at 2.9~eV (Figure \ref{fig:exc_weight}d) are similar. 
\begin{figure}[!htb]
\centering
\includegraphics[width=9cm]{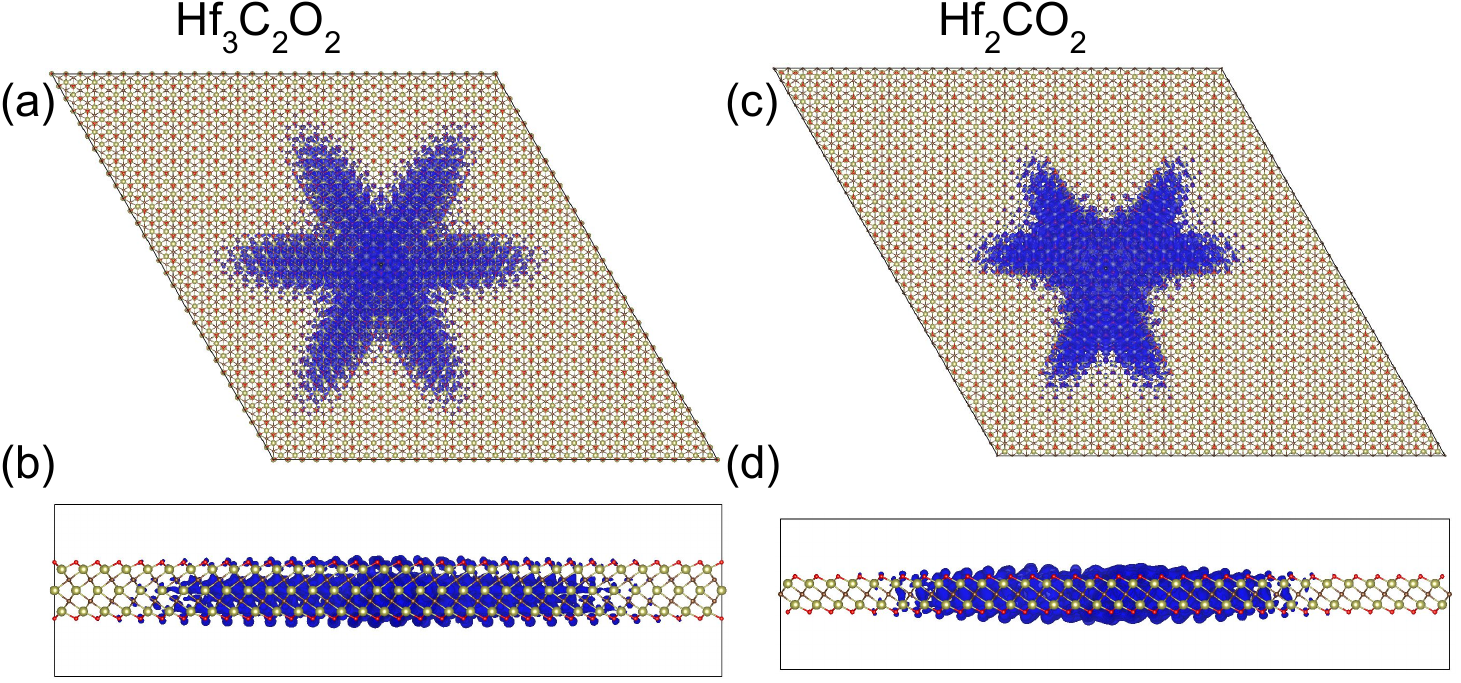}
\caption{\label{fig:isosurface_one} Isosurface (with value of 3$\times10^{-9}$) for the real space electron probability density of the first bright exciton in (a,b) Hf$_3$C$_2$O$_2$ and (c,d) Hf$_2$CO$_2$. The hole is located above a carbon atom, 1 \AA~ away in the z-direction in both materials.
}
\end{figure}

We also explored the spatial distribution of excitons by fixing the position of the hole within the material and examining the probability of finding the electron in different spatial locations. 
This approach provided valuable insights into the spatial characteristics of excitonic states.\cite{kolos2021giant,kolos2022large}
By fixing the hole's position and studying the electron's probability distribution, we could distinguish between localized and delocalized excitonic states within the material.\cite{chernikov2014exciton,cudazzo2016exciton} 
We focused here on the first bright exciton states in Hf$_3$C$_2$O$_2$ (at 0.9~eV) and Hf$_2$CO$_2$ (at 2.7~eV). 
These low-energy excitons are, in fact, consisting of four degenerated BSE states in both materials. Nevertheless, in Hf$_3$C$_2$O$_2$, two are dominant (by the strength), with the other two dark (strength is negligible) and not influencing spatial distribution or optical spectrum. Hf$_2$CO$_2$ is a slightly different situation where from $\lambda$ = 9-12 BSE eigenstates is just $\lambda$ = 11 bright and others more or less dark. 
For the real-space plot of Figure \ref{fig:isosurface_one}, we fixed the position of the hole located at the top of the C atom at a distance of 1 {\AA} in the z-direction for both materials.

Our observations show that the wave functions associated with individual BSE eingenstates exhibit a notable deviation from the intrinsic symmetry characteristics of the host materials. 
Specifically, the electron probability densities within these BSE eingenstates do not conform to a rotationally symmetric distribution. 
This asymmetry in electron distribution is an intriguing aspect, as it contrasts with the expected behavior based on the material's inherent properties. 
However, an interesting phenomenon arises when we consider the cumulative effect of the electron probability densities. 
By aggregating the densities of all parts of excitons in Hf$_3$C$_2$O$_2$ and Hf$_2$CO$_2$, an alignment with the material's native threefold rotational symmetry emerges, as can be seen in Fig. \ref{fig:exc_weight} subplots and Fig. \ref{fig:isosurface_one}. This suggests that while individual BSE states might exhibit asymmetrical properties, their collective behavior can reflect and restore the material's symmetry.  
Further, the exciton responsible for the first dominant peak in Hf$_3$C$_2$O$_2$ spectrum is very delocalized, and we were forced to use at least 30$\times$30$\times$1 k-mesh grid.
Therefore, we needed a supercell of 30 real unit cells to ensure the wave function fades out on the border of the supercell (Figure \ref{fig:isosurface_one}). 
For smaller k-grids/supercells, the artificial confinement of the exciton led to a biased optical gap and untruly increased the predicted binding energy of the exciton (Figure \ref{fig:convergence_gap}).
On the other hand, the first bright exciton in Hf$_2$CO$_2$ was more localized (cf. isosurfaces of Figure \ref{fig:isosurface_one}).
In our investigation into the variance of the first exciton's spatial extent, we gain insights into the differing rates of convergence of the optical band gaps in Hf$_3$C$_2$O$_2$ and Hf$_2$CO$_2$ relative to the number of k-points. 
Building upon the findings discussed in our preceding work,\cite{Kolos2022} we understand that an exciton with a widespread distribution in real space necessitates a higher number of k-points for accurate characterization. 
Consequently, the optical band gaps in Hf$_3$C$_2$O$_2$ exhibit a slower convergence concerning the number of k-points compared to Hf$_2$CO$_2$ (Fig. \ref{fig:convergence_gap}). 
This slower convergence is attributed to the broader spatial distribution of the first bright exciton in Hf$_3$C$_2$O$_2$, as opposed to the more localized nature of the exciton in Hf$_2$CO$_2$.

In examining the symmetry of the first bound exciton in both materials, the influence of spin-orbit coupling (SOC) on the degeneracy (splitting of bands) and the respective symmetry groups D$_{3h}$ for Hf$_3$C$_2$O$_2$ and D$_{3d}$ for Hf$_2$CO$_2$, it is evident that these excitons align with the E-type irreducible representations because these are associated with degeneracy. Specifically, for Hf$_3$C$_2$O$_2$, the first bright exciton can be associated with the E$^\prime$ and E$^{\prime \prime}$ representations. Meanwhile, Hf$_2$CO$_2$ first bright exciton can correspond to the E$_g$ and E$_u$ representations. The real-space exciton projection for Hf$_3$C$_2$O$_2$ (Figure \ref{fig:isosurface_one} (a) and (b)) exhibits vertical plane symmetry, leading to the assignment of the E$^\prime$ group. In contrast, the gerade centrosymmetric nature observed in the Hf$_2$CO$_2$ exciton projection (Figure \ref{fig:isosurface_one} (c) and (d)) suggests its alignment with the E$_g$ group.
\begin{figure}[!htb]
  \centering
\includegraphics[width=8.35cm]{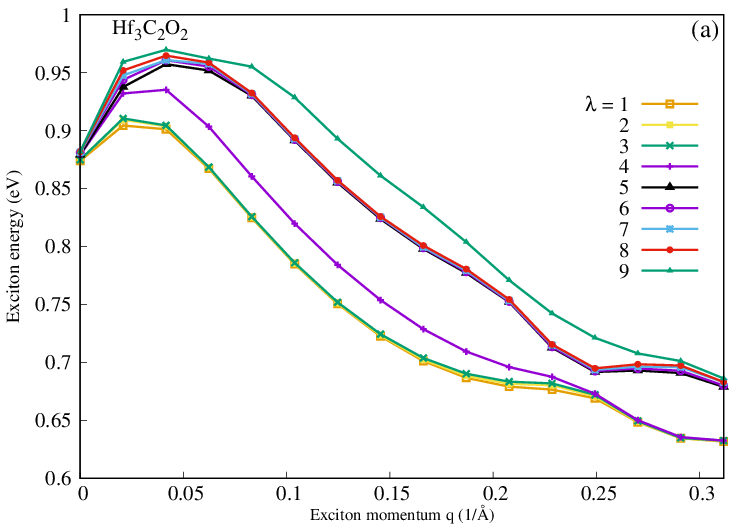}
\includegraphics[width=8.35cm]{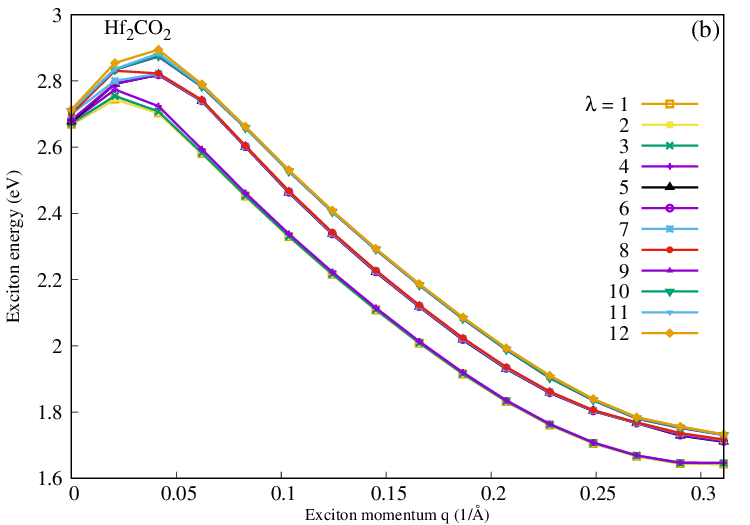}
\caption{\label{fig:finite_exciton}Excitation energy of the (a) Hf$_3$C$_2$O$_2$ and (b) Hf$_2$CO$_2$ MXene with exciton momentum {$\textbf{q}$} along {$\Gamma$} to M path.}
\end{figure}

Concerning the indirect nature of both investigated materials (see indirect bands in Figure \ref{fig:bandstructure}), we finally performed excitonic BSE calculations for the indirect exciton momentum ($\textbf{q}$ $\neq$ 0). 
In indirect materials, it leads to a decrease in excitation energies, or, in extreme cases of excitonic insulator candidates, it can lead to negative excitation energies, i.e., the situation where the binding energy of excitons overcomes the fundamental gap.\cite{kumar2023oxygen} 
Here, following the band structure of Figure \ref{fig:bandstructure}, we were focused on exciton momentum along {$\Gamma$}M direction. 
The calculations reveal that the exciton energies are (after a slight increase) significantly decreasing, approaching a minimal value to the M point - see Figure \ref{fig:finite_exciton}. 
The non-extrapolated excitation energy of 0.88 eV in Hf$_3$C$_2$O$_2$ was reduced to 0.68 eV (ca. to 78\% of the original value). 
In the Hf$_2$CO$_2$ case, the reduction was extreme, from 2.70 eV to 1.73 eV (ca. to 64\% of the original value).

Our last analysis lies in the guess of exciton behavior in time (Table \ref{tab_radiative_lifetime}). 
Using Eqn. \ref{radiative_lifetime}, we calculated the radiative lifetime $ \tau_{\lambda}^{0}$ (free from temperature dependency) for the first bright excitons in the Hf$_3$C$_2$O$_2$ and Hf$_2$CO$_2$ MXenes, 0.22 ps and 0.02 ps, respectively. These lifetimes indicate rapid radiative recombination processes, showing that light emission from these excitonic states is highly efficient. 
Values of $ \tau_{\lambda}^{0}$ for Hf$_3$C$_2$O$_2$ are comparable with best-known monolayer transition metal dichalcogenides (TMDs) because the
radiative exciton lifetime in MoS$_2$, MoSe$_2$, WS$_2$, or WSe$_2$ is of order 0.1-1 ps.\cite{palummo2015exciton} 
On the other hand, the first bright exciton in Hf$_2$CO$_2$ has a shorter lifetime of one order of magnitude, i.e., comparable with exciton lifetimes in two‐dimensional group‐III nitrides as BN, AlN, GaN, and InN, which are of 0.01 ps order.\cite{Prete2020}
\begin{table}[htb]
\setlength\tabcolsep{2pt} 
\caption{\label{tab_radiative_lifetime}{} Zero-temperature $\tau_{\lambda}^{0}$ and average room-temperature $\langle \tau^{RT}_{\lambda} \rangle$ radiative lifetimes of the first bright excitons (all degenerated BSE eigenstates $\lambda$ are included) in Hf$_3$C$_2$O$_2$ and Hf$_2$CO$_2$, where $ M_{\lambda}$ is the mass of exciton and $E_{\lambda}$ is the extrapolated energy of the first bright excitons.} 
\begin{tabular}{ccc ccc}
\hline
\hline
& {Hf\textsubscript{3}C\textsubscript{2}O\textsubscript{2}} & {Hf\textsubscript{2}CO\textsubscript{2}}\\
\hline
$E_{\lambda}$ (eV) & 0.91 & 2.72\\
 {$\mu^2_\lambda$}/{$A_{uc}$} 
 & 0.04 & 0.12\\
 $ M_{\lambda}$ (a.u.) & 2.22 & 0.92 \\
 {$ \tau_{\lambda}^{0}$} (ps) & 0.22 & 0.02\\
 $\langle \tau^{RT}_{\lambda} \rangle$ (ns) & 11.44 & 0.06\\
\hline
\hline
\end{tabular}
\end{table}
For computing the average radiative lifetime $\langle \tau^{RT}_{\lambda} \rangle$ of exciton at temperature $T$, it was necessary to calculate the exciton mass $M_{\lambda}$, 2.22~a.u. and 0.92~a.u. for the Hf$_3$C$_2$O$_2$ and Hf$_2$CO$_2$ monolayers, where we sum effective masses of electron and hole\cite{spataru2005theory} in M point of BZ. 
At room temperature (300~K) using Eqn. \ref{avg_radiative_lifetime}, the average radiative lifetimes $\langle \tau^{RT}_{\lambda} \rangle$ of Hf$_3$C$_2$O$_2$ and Hf$_2$CO$_2$ are 11.44~ns and 0.06~ns. 
While in Mo- and W- based TMDs, the first bright exciton has $\langle \tau^{RT}_{\lambda} \rangle$ of tenths of ns, it is much longer in Hf$_3$C$_2$O$_2$. 
This behavior mostly comes from Hf$_3$C$_2$O$_2$ first bright exciton energy (0.9~eV) in the infrared region, while TMDs first bright exciton energies (1.7-2.0~eV)\cite{palummo2015exciton} are in the visible region (see also $E_{\lambda}^2$ in Equation 5). 
We note that the short average radiative lifetime of the first bright exciton in Hf$_2$CO$_2$ is rather similar to excitonic lifetimes in monolayer BN, AlN, and GaN (tens of ps).\cite{Prete2020} 
Linear dependence of the average radiative lifetimes with respect to the temperature (Eqn. \ref{avg_radiative_lifetime}) has a slope of 38.13~ps/K and 0.19~ps/K for Hf$_3$C$_2$O$_2$ and Hf$_2$CO$_2$, respectively.  
The Hf$_3$C$_2$O$_2$ value is comparable to that of the beforementioned TMDs.\cite{palummo2015exciton}



\section{Conclusions}
\label{conclusions}
In conclusion, motivated by recent discoveries of Hf-based MXenes, we examined the geometric and optoelectronic properties with many-body methods. We presented a detailed study of the fundamental properties following linear optics of new 2D materials, Hf-based MXenes, Hf$_3$C$_2$O$_2$ and Hf$_2$CO$_2$. Both semiconducting Hf$_3$C$_2$O$_2$ and Hf$_2$CO$_2$ MXenes have an indirect quasiparticle gap of 1.08~eV and 2.18~eV, respectively. 
Both Hf$_3$C$_2$O$_2$ and Hf$_2$CO$_2$ MXenes showed strong exciton binding energies of 0.51~eV and 0.76~eV, respectively. 
The excitation energies of direct excitons can be significantly decreased by considering indirect excitons with exciton momentum {$\textbf{q}=\Gamma M$}. 
The observed 4{\%} and 13{\%} absorptance in the infrared region and the remarkable 17{\%} peak at 4.8 eV in the near UV region highlight the rich excitonic landscape of MXene Hf$_3$C$_2$O$_2$. 
Additionally, MXene Hf$_2$CO$_2$ exhibited a significant 28{\%} absorptance peak in the visible spectra. 
The emission properties of the Hf-based MXenes are illustrated by the radiative lifetimes of the lowest-energy excitons. 
We observed almost three orders' lifetime difference between three-Hf-layered and two-Hf-layered materials at room temperature: very long-lived ($\sim$ ten nanoseconds) excitons in Hf$_3$C$_2$O$_2$ (longer than in transition metal dichalcogenides) vs. short living ($\sim$ tens of picoseconds) in Hf$_2$CO$_2$.
\section*{Supplementary Material}
See \textcolor{blue!80!black}{supplementary material} for the Hf-based MXenes structures used in the reported computations, optimization of the conformers, and convergence of the necessary parameters used in DFT, GW, and BSE calculations. 
\section*{Acknowledgments}
This work was supported by the Czech Science Foundation (21-28709S) and the University of Ostrava (No. SGS06/PrF/2023, SGS04/PrF/2024). 
This work was partially carried out with financial support from SERB and DST, India, with grant number SERB-MATRICS MTR/2021/000017.
The calculations were performed at IT4Innovations National Supercomputing Center (e-INFRA CZ, ID:90140). 
We also acknowledge the National SuperComputing Mission (NSM) for providing computing resources of ’PARAM Smriti' and 'PARAM Shivay' at NABI, Mohali, and IIT BHU, India, respectively, for the computations.
\bibliography{refs}

\begin{thebibliography}{52}
\bibitem [{yam()}]{yambo_mu}%
  \BibitemOpen
  \href@noop {} {}\bibinfo {note} {'\texttt{Strengths}' output printed out by a
  post-processing tool (YPP) included in the Yambo code (version 5.1.0) in eV,
  must be converted to $\mu^2_\lambda$ (in SI units):
  $\mu^2_\lambda$=$\frac{\epsilon_0 V}{e^2}$\texttt{Strengths}, where
  $\epsilon_0$ is the vacuum permittivity, $V$ is the volume of the
  computational cell, and $e$ is the elementary charge.}\BibitemShut {Stop}%
\end{thebibliography}

\end{document}